\documentclass[conference]{IEEEtran}

\usepackage[T1]{fontenc}
\usepackage{color}

\usepackage{cite}

\ifCLASSINFOpdf
  \usepackage[pdftex]{graphicx}
\else
  \usepackage[dvips]{graphicx}
\fi

\usepackage{amsmath,amssymb}
\def\braket#1{\mathinner{\langle{#1}\rangle}}

\def\Ket#1{\left|#1\right\rangle}

\usepackage{array}

\usepackage{url}

\makeatletter
\newcommand*{\Rom}[1]{\expandafter\@slowromancap\romannumeral #1@}
\newcommand*{\rom}[1]{\expandafter\romannumeral #1}
\makeatother

\DeclareMathOperator*{\argmin}{\text{argmin}}

\usepackage[normalem]{ulem}

\begin{document}

%
\title{Quantum Backscatter Communication: A New Paradigm}
%
%
%
\author{\IEEEauthorblockN{
Roberto Di Candia\IEEEauthorrefmark{1},
Riku J\"{a}ntti\IEEEauthorrefmark{2},
Ruifeng Duan\IEEEauthorrefmark{2},
Jari Lietz\'{e}n\IEEEauthorrefmark{2},
Hany Khalifa\IEEEauthorrefmark{2}, and
Kalle Ruttik\IEEEauthorrefmark{2}
}
\IEEEauthorblockA{\IEEEauthorrefmark{1}Dahlem Center for Complex Quantum Systems, Freie Universit\"{a}t Berlin, 14195 Berlin, Germany.\\
Email: rob.dicandia@gmail.com}
\IEEEauthorblockA{\IEEEauthorrefmark{2}Department of Communications and Networking, Aalto University, Espoo, 02150 Finland.\\ Email: \{riku.jantti; ruifeng.duan; jari.lietzen; hany.khalifa; kalle.ruttik\}@aalto.fi}
}


%



\maketitle
\begin{abstract}
In this paper, we propose a novel quantum backscatter communications (QBC) protocol, inspired by the quantum illumination (QI) concept. In the QBC paradigm, the transmitter generates entangled photon pair. The signal photon is transmitted and the idler photon is kept at the receiver. The tag antenna communicates by performing the pulse amplitude modulation (PAM), binary phase shift keying (BPSK) or quadratic phase shift keying (QPSK) on the signal impinging at the antenna. Using the sum-frequency-generation receiver, our QBC protocol achieves a 6 dB error exponent gain for PAM and BPSK, and 3 dB gain for QPSK over its classical counterpart. Finally, we discuss the QI-enhanced secure backscatter communication.
\end{abstract}


%


\section{Introduction}

Backscatter of radio waves is the subject of active study since the development of radar in the 1930s, and the use of it for communications since 1948 \cite{Stockman1948}. Backscatter communication (BC) is widely used in radio frequency (RF) identification tags, and it bears close resemblance with the radar. Quantum radar \cite{Allen2008patent} is a remote-sensing method based on quantum entanglement. Quantum illumination (QI) was introduced by S. Lloyd in 2008 with the idea of using entangled photons to increase the success probability of detecting a low-reflectivity object in a noisy and lossy environment~\cite{Lloyd2008, Tan08}. The application in the microwave regime was proposed afterwards, and it paved the way to a prototype of quantum radar~\cite{Barzanjeh2015}. QI has also been utilized for quantum key exchanged in optical communication systems \cite{Shapiro2014}. In this paper, we propose to use QI to enhance  backscatter communications. The proposed Quantum Backscatter Communications (QBC) bears close resemblance to the quantum radar in a manner similar to BC being closely related to the classical radar. Our recent work paper \cite{Jantti2017} has proposed to construct pre-coder beam-splitters and receiver beamforming beam-splitters such that the orthogonal eigen-channels can be accessed using QBC. This paper aims at describing the QBC concept and analyzing its performance in terms of bit error rate (BER).


In the proposed QBC paradigm, the reader antennas are pointed toward a tag antenna that communicates by performing the pulse amplitude modulation (PAM), binary phase shift keying (BPSK) or quadratic phase shift keying (QPSK) on the signal. The receiver is then able to discriminate the states of the idler-signal system after the corresponding phase and amplitude modulation are performed by the tag. We compare the classical architecture, consisting in illuminating the antenna with classical light or microwave signal and performing heterodyne detection, with the quantum architecture using  Gaussian entangled states as resources. We quantify the quality of the performance in a Bayesian setting by seeking the best scaling of the error probability (EP) averaged over the {\rm a priori} probabilities of each symbol. In the PAM and BPSK cases, the quantum setting allows for up to $6$ dB improvement in the EP exponent (EPE)~\cite{Tan08}. This can be achieved by a slight modification of the Zhuang receiver (interchangeably called SFG-RX) proposed in~\cite{Zhuang2017}, based on a sum-frequency-generation (SFG) circuit, parametric amplifiers (PAs), photon-counter and a feedback loop. Alternatively, a simpler circuit based only on parametric amplification and photon-counting can achieve a $3$ dB gain~\cite{Guha2009, Sanz17}, which has been implemented in a laboratory environment in the optical regime~\cite{Zhang2015}. We further show that only SFG-RX achieve a $3$ dB gain in the QPSK case. At the end, we argue that the BPSK and QPSK schemes are useful for quantum cryptography, allowing for secure communication between the antenna and the RX.



\subsection{States, Observables, and Quantum Harmonic Oscillator}
In quantum mechanics, the \emph{state of a system} can be represented by column vector $|\psi\rangle\in \mathbb{C}^{\infty}$, normalized as ${\langle \psi| \psi\rangle=1}$.~\footnote{We use the bra-ket notation. The scalar product between two states $|\phi\rangle$ and $|\psi\rangle$ is denoted by $\langle \phi|\psi\rangle\equiv \sum_{i}\phi_i^*\psi_i$, where $^*$ denotes complex conjugate of a number. The operators $\hat A\in\mathbb{C}^{\infty\times \infty}$ apply only on the right so that $\langle\phi|\hat A|\psi\rangle=\sum_{i}\phi_i^*(\hat A\psi)_i=[\sum_{i}\psi_i^*(\hat A^\dag \phi)_i]^*=\langle \psi|\hat A^\dag |\phi\rangle^*$, where $\dagger$ is an adjoint operator.} \emph{Observables} are Hermitian~\footnote{Despite non-Hermitian operators, i.e., $\hat A\not=\hat A^\dag$, do not represent measurable quantities, it is useful to formally define their expectation values on the state $|\psi\rangle$ as $\langle\hat A\rangle_{\psi}\equiv \langle\psi| \hat A|\psi\rangle \in\mathbb{C}$.} operators $\hat O \in \mathbb{C}^{\infty\times \infty}$. The expectation value of the observable $\hat O$ on the state $|\psi\rangle$ is defined as $\langle\hat O\rangle_{\psi}\equiv\langle \psi|\hat O | \psi\rangle\in \mathbb{R}$.  The \emph{Hamiltonian of a system} is an operator $H\in\mathbb{C}^{\infty\times \infty}$, which rules the evolution of the state of the system via the Schr\"odinger's equation: $i\hbar\partial_t |\psi(t)\rangle= H|\psi(t)\rangle$, where $\hbar$ denotes the reduced Planck constant and $i=\sqrt{-1}$. These concepts can be easily generalize to bipartite systems by using the tensor product formalism. The joint state of two separate systems is represented by a vector $|\psi\rangle_{AB}\in {\mathbb{C}^{\infty}\otimes \mathbb{C}^{\infty}}$, where $\otimes$ denotes the tensor product and ${}_{AB}\langle \psi| \psi\rangle_{AB}=1$. Observables in bipartite systems are Hermitian operators $\hat O_{AB}\in\mathbb{C}^{\infty\times \infty}\otimes\mathbb{C}^{\infty\times \infty}$.~\footnote{Observables of individual systems are operators $\hat O_A\otimes \mathbb{I}$ and $\mathbb{I}\otimes \hat O_B$, and are usually denoted simply as $\hat O_A$ and $\hat O_B$. Here $\mathbb{I}$ is the identity operator.} A bipartite state $|\psi\rangle_{AB}$ is said to be {\it entangled} if it can not be decomposed as $|\psi\rangle_{AB}=|\phi\rangle_A\otimes |\chi\rangle_B$. Entangled states show correlations between outcomes of measurements on the individual systems which are not reproducible in the classical systems, and they will be a key resources for the results in this paper.

In the low photon number regime, if the thermal fluctuation are negligible, the electromagnetic field behaves according to quantum electrodynamic theory. The free Hamiltonian of quantized electromagnetic field has the same form as the Hamiltonian of quantum harmonic oscillator  $H=\hbar\omega\left(\hat{a}^\dagger\hat{a}+1/2\right)$ with the annihilation operator $\hat {a}$ and the creation operator $\hat {a}^\dag$. Their action on the basis defined by the eigenvectors $\{|n\rangle \}_{n=0}^\infty$ of the Hamiltonian $H$ is $\hat {a}|n\rangle=\sqrt{n}|n-1\rangle$ and $\hat {a}^\dag |n\rangle=\sqrt{n+1}|n+1\rangle$. Moreover, the commutation relation $[\hat a, \hat a^\dag]=1$ hold. The eigenstates of the operator $\hat a$ are referred to as coherent states, and we refer to them as classical states, because the statistics of their measurements resembles the one of the classical signals. The canonical position and momentum-like operators are given by $\hat x\equiv(\hat {a}+\hat {a}^{\dag})/\sqrt{2}$ and $\hat p\equiv-i(\hat {a}-\hat {a}^{\dag})/\sqrt{2}$ respectively.

\subsection{Backscatter Communication}
The BC consists in sending a signal to a tag, which chooses a symbol $(\sqrt{\eta},\phi)$ belonging to alphabet $\mathcal{A}$ that defines a particular BC scheme. In this work we consider the following BC schemes using PAM, BPSK, and QPSK modulation techniques:
\begin{IEEEeqnarray*}{ll}
    \mathcal{A}_\mathsf{PAM} &=\left\{(\sqrt{\eta_1},0),(\sqrt{\eta_2},0)\right\}, \\
\mathcal{A}_\mathsf{BPSK} &=\left\{(\sqrt{\eta},0),(\sqrt{\eta},\pi)\right\}, \\
\mathcal{A}_\mathsf{QPSK} & =\left\{(\sqrt{\eta},0),(\sqrt{\eta},\pi/2),(\sqrt{\eta},\pi),(\sqrt{\eta},3\pi/2)\right\}.
\end{IEEEeqnarray*}

In this framework, the QI setup previously studied in the literature corresponds to the PAM case with $\eta_1=0$. In this paper, we consider the symmetric case, where all the symbols are chosen with equal {\it a priori} probability, and the performance of the BC scheme is quantified by the EP, defined as
\begin{equation}
    P_{\mathsf{err}}=\frac{1}{|\mathcal{A}|}\sum_{(\eta,\phi)\in\mathcal{A}}\text{Pr}\,[\large\bar{H}_{(\eta,\phi)}|H_{(\eta,\phi)}],
\end{equation}
where $|\mathcal{A}|$ is the size of $\mathcal{A}$ and $\text{Pr}\,[\large\bar{H}|H]$ is the probability that, given the hypothesis $H$, we wrongly declare a different hypothesis. The EP $P_{\mathsf{err}}$ decays exponentially in both the classical and in the quantum case. We will show a quantum-RX decaying at a higher rate with respect the classical case.

The rest of this paper is organized as follows. In Section~\ref{sec:sys_model}, we describe a bistatic QI-enhanced quantum backscatter system. Section~\ref{sec:Qreceivers} introduces the quantum receivers and discusses quantum protocols, and we then compare the performance of classical and quantum receivers. We discuss in Sec.~\ref{Sec:security} the possibility of using QI to enhance the secure backscatter communication. Finally, Section~IV concludes the paper.

\begin{figure}[!t]
	\centering
	\includegraphics[width=0.9\columnwidth]{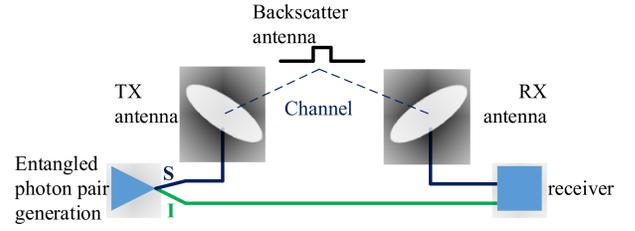}	
	\caption{An illustration of bistatic QI-enhanced QBC systems.}
	\label{fig:sys}
\end{figure}

\section{System model}\label{sec:sys_model}

The quantum backscatter system shown in Fig.~\ref{fig:sys} enhances the feature of a classical QBC using entangled photons. In the classical case, the transmitter (TX) transmits an unmodulated carrier backscattered from the tag to the receiver (RX). The carrier is modeled in quantum mechanics as a coherent state $|\alpha\rangle_S=e^{-|\alpha|^2/2}\sum_{n=0}^\infty \sqrt{{\alpha^n}/{n!}}\,|n\rangle$, where $\alpha$ is a parameter related to the mean number of signal photons.

In the QI setup, the entangled signal-idler (S-I) photon pairs are first generated at the TX. The S-photon is transmitted and backscattered from a tag antenna. The idler is kept at the RX  to be measured jointly with the backscattered signal. The RX uses both the received S-photon from the RX antenna and the I-photon. The system in Fig.~\ref{fig:sys} is bistatic\footnote{We assume that the I-photons are available for the RX without losses. In mono-static case, the idler photon is directly available at the RX. In the bistatic case, it would need to be transmitted over a cable to the RX. In practice this transmission will cause losses which will reduce the system gain.}, meaning that the transmitter and the RX are separated in space. We consider a source able to continuously generate S-I photon pairs in the radio frequency regime in a two-mode squeezed state (TMSS)
\begin{equation*}
    \Ket{\psi}_{SI}=\sum_{n=0}^\infty \sqrt{{N_S^n}/{(N_S+1)^{n+1}}\, \Ket{n}_S\Ket{n}_I},
\end{equation*}
where $N_S$ is the average number of photons of both the signal and the idler \cite{Tan08,Guha2009,Sanz17,Zhuang2017}. The joint probability distribution of the quadratures of the TMSS is a Gaussian with zero mean value, hence the state is well defined by its covariance matrix. Indeed, if $\hat a_S$ and $\hat a_I$ represent the modes of the signal and the idler, respectively, then we have that $\braket{\hat{a}_S^\dagger\hat{a}_S}=\braket{\hat{a}_I^\dagger\hat{a}_I}=N_S$,  $\braket{\hat{a}_S\hat{a}_I}=\sqrt{N_S(N_S+1)}$ and $\braket{\hat{a}_S^\dag\hat{a}_I}=0$.

We model the quantum channel with a low-reflectivity beam splitter, whose inputs are the signal and a thermal state modeling the effect of the environment. This is described with the unitary operator
\begin{equation*}
    B_{\eta,\phi}=\exp{\left[\sin^{-1}(\sqrt{\eta})\left(\hat {a}_S^\dag \hat{a}_Ze^{-i\phi}-\hat {a}_S \hat {a}_Z^\dag e^{i\phi}\right)\right]},
\end{equation*}
where $\phi$ is the phase shift, and $\eta$ is the round-trip transmissivity (RTT) of the channel. Here, the environmental thermal mode $\hat a_Z$ is a Gaussian mode. The number of thermal photons $N_Z=(e^{\hbar \omega/k_BT}-1)^{-1}$, where $T$ is the environment temperature and $k_B$ the Bolzmann constant. Both the impact of the propagation path loss and the tag antenna are included in the effective parameters $\eta\ll 1$ and $\phi$. The input-output relations of the beam splitter read
\begin{IEEEeqnarray}{lllll}\label{rec}
    \hat {a}_R & \equiv & {B}_{\eta,\phi}^\dag \hat{a}_S {B}_{\eta,\phi} & = & \sqrt{\eta} e^{-i\phi}\hat {a}_S +\sqrt{1-\eta}\hat {a}_Z, \\
    \hat {a}_Y & \equiv & {B}_{\eta,\phi}^\dag \hat{a}_Z {B}_{\eta,\phi} & = & -\sqrt{\eta} e^{i\phi}\hat {a}_Z +\sqrt{1-\eta}\hat {a}_S,
\end{IEEEeqnarray}
where $\hat {a}_R$ is the received mode, and $\hat {a}_Y$ corresponds to modes that are not received and can thus be ignored.

Depending on the transmit and receive antenna gains $G_t$ and $G_r$ of the reader, distance from the transmitter to the tag $R_t$, the communication frequency $\omega$, and the distance from tag to RX $R_r$, the RTT $\eta$ can be represented as:
\begin{equation*}
    \eta=\frac{G_rG_t\mathsf{c}^2\sigma_Q}{(16\pi \omega^2 R_t^2R_r^2)}.
\end{equation*}
The parameter  $\sigma_Q={\braket{\hat{I}_s}}/{\braket{\hat{I}_i}}$ is the Quantum Radar Cross Section (QRCS) \cite{Liu2014}, where $\braket{\hat{I}_s}$ denotes the intensity measured by a detector after a photon is reflected by atoms on the target surface, the tag antenna in our case, and $\braket{\hat{I}_i}$ is the incident intensity calculated assuming the target to act as a photon detector. The phase shift of the channel $\phi$ depends on the communication distance $R=R_t+R_r$ and the phase shift caused by the tag $\phi=2\pi R /\mathsf{c} + \varphi$. A large number of mode pairs $M$ are needed in order to perform the QSD at the RX. The available number of mode pairs $M=W T_s$ depends on the phase matching bandwidth $W$ and the tag symbol duration $T_s$ assumed to be small compared to the channel coherence time.

\section{Performance Analysis}\label{sec:Qreceivers}
\subsection{Classical receiver}
In the base-line classical case (referred to as C-RX), i.e. without using QI technique, the carrier is a Gaussian mode in a coherent state with an average number of photons $N_S$. Heterodyne detection is then performed on each of the $M$ copies of the received mode $\hat a_R$ defined in \eqref{rec}. Heterodyne detection is modeled in the quantum formalism as a $50:50$ beam splitter with outputs $\hat a_1=({\hat a_R+\hat a_V})/{\sqrt{2}}$ and $\hat a_2=({\hat a_R-\hat a_V})/{\sqrt{2}}$, where $\hat a_V$ is a complex Gaussian noise with variances $\langle \hat x_V^2\rangle = \langle \hat p_V^2\rangle={1}/{2}$. Let us introduce the complex envelope $\hat S\equiv \left(\hat x_1+i\hat p_2\right)/{\sqrt{N_S}}$ with mean value $\langle\hat S\rangle=\sqrt{\eta}e^{-i\phi}$, given that $\langle \hat x_1\rangle=\sqrt{\eta N_S}\cos(\phi)$ and $\langle \hat p_2\rangle=-\sqrt{\eta N_S}\sin(\phi)$. If we measure $M\gg1$ times both $\hat x_1$ and $\hat p_2$, we can estimate the mean value of the complex envelope with the sample mean $\bar S\equiv \sum_{i=1}^M S_i/{M}$. Finally, we declare the symbol $\{\sqrt{\tilde \eta},\tilde \phi\}=\argmin_{\{\sqrt{\tau},\varphi\}\in \mathcal{A}_L}\left|\bar S-\sqrt{\tau}e^{-i\varphi}\right|$, and the protocol succeeds if $\{\sqrt{\tilde \eta},\tilde \phi\}=\{\sqrt{\eta},\phi\}$. In the classical setup, regardless the BC scheme used, it is well-known that the EP $P^c_{\mathsf{err}}$ is lower-bounded by 
\begin{equation}
    P^c_{\mathsf{err}}\geq \frac{1}{2|\mathcal{A}|}\text{\rm erfc}\left(\sqrt{\frac{\min_{{\tilde k}\not= k}d_{k{\tilde k}}^2N_SM}{4N_Z}}\right)\sim e^{-\frac{\min_{{\tilde k}\not= k}d_{k{\tilde k}}^2N_SM}{4N_Z}},
    \label{eq:Perrlbclassical}
\end{equation}
where $d_{k{\tilde k}}\equiv|\sqrt{\eta_k}e^{-i\phi_k}-\sqrt{\eta_{\tilde k}}e^{-i\phi_{\tilde k}}|$ and the minimum is taken over the elements of $\mathcal{A}$. This provides the asymptotic behaviour of the classical EP. For the considered BC schemes we have that: {$|\mathcal{A}_\mathsf{PAM}|=2$} and {$d_{k{\tilde k}}^2=|\sqrt{\eta_2}-\sqrt{\eta_1}|^2\equiv d_{\mathsf{PAM}}^2$} for PAM; {$|\mathcal{A}_\mathsf{BPSK}|=2$} and {$d_{k{\tilde k}}^2=4\eta\equiv d_{\mathsf{BPSK}}^2$} for BPSK; $|\mathcal{A}_\mathsf{QPSK}|=4$ and $\min_{{\tilde k}\not= k}d_{k{\tilde k}}^2=2\eta\equiv d_{\mathsf{QPSK}}^2$ for QPSK.

\subsection{Quantum receiver}

In the quantum case, different symbols correspond to different quantum states at the RX. Therefore, the task reduces to find a measurement discriminating with the least number of measurements, between the quantum states which are the possible outputs of the considered QBC scheme. A practical requirement consists in finding an experimentally feasible circuit achieving the optimal measurement. The optimal decision rule for discriminating between two equally likely quantum states $\rho_0$ and $\rho_1$ was found by Helstrom~\cite{Helstrom1976}. It consists in measuring the operators ${E_0,E_1}$, with $E_0+E_1=\mathbb{I}$, where $E_1$ is the projection on the range of the positive part of $\rho_1-\rho_0$. The corresponding optimal EP is $P_H=(1-\| \rho_0-\rho_1\|_1)/2$. In the multiple copies case, where we need to discriminate between the two states $\rho_0^{\otimes M}$ and $\rho_1^{\otimes M}$, with $M\gg1$, the computation of the optimal probability and the corresponding measurement can be substantially challenging. An upper bound on the EP is provided by the {\it quantum Chernoff bound} (QCB)~\cite{Verstraete07}, stating that $P_H\leq e^{-M\xi_{QCB}}$, where $\xi_{QCB}=-\log \left(\min_{0\leq s\leq 1}\text{Tr}\,(\rho_0^s\rho_1^{1-s})\right)$. This bound is asymptotically tight, i.e. $P_H \sim e^{-M\xi_{QCB}}$ for $M\gg1$.

In the QI setup, the QCB has been computed for PAM case in~\cite{Tan08}, showing a  $6$~dB gain over the best classical strategy. Recently, a measurement saturating the Helstrom EP has been obtained in \cite{Zhuang2017}. It applies an SFG to the modes at the RX, allowing to map the problem to the discrimination between two coherent states, where the Dolinar RX is known to be optimal for this task~\cite{Dolinar1973}. A suboptimal RX, achieving a $3$~dB gain and consisting in a PA and photon-counting, has been proposed in \cite{Guha2009} and implemented in \cite{Zhang2015}. The performance loss is accompanied with a benefit in the experimental feasibility, as the latter RX involves only two-mode interactions in contrast to three-mode interactions needed in the SFG-RX. We show how these RXs achieve a gain in QBCs.

\emph{PA-receiver}: It consists in the measurement of the observable $\hat O_{PA}=\hat a_I \hat a_R+\hat a_I^\dag \hat a_R^\dag$, with mean value $\langle \hat O_{PA}\rangle=\sqrt{\eta N_S(N_S+1)}\cos(\phi)$ and variance\footnote{It has been approximated in the $\eta\ll 1, N_S\ll 1, N_Z\gg 1$ limits.} $\langle O_{PA}^2 \rangle -\langle \hat O_{PA}\rangle^2\approx N_Z$. This is implemented with the help of a PA, whose input-output relations are 
\begin{equation}
    \hat c = \sqrt{G} \hat a_I+\sqrt{G-1}\hat a_R^\dag \quad \text{and}\quad      \hat d = \sqrt{G} \hat a_R+\sqrt{G-1}\hat a_I^\dag.
\end{equation}
If we choose $G=1+\varepsilon^2$, with $N_S/N_Z\ll \varepsilon^2\ll 1/N_Z$~\cite{Guha2009}, then the photon-number operator approximates $\hat O_{PA}$:
\begin{equation}
    \hat c^\dag \hat c = G\,\hat a^\dag_I \hat a_I+(G-1)\,\hat a_R \hat a_R^\dag +\sqrt{G(G-1)} \,\hat O_{PA} \approx \varepsilon\, \hat O_{PA}, 
\end{equation}
where we have used that $\varepsilon^2\ll1/N_Z$ and $\langle \hat a_I^\dag \hat a_I\rangle=N_S\ll1$ in order to conclude that $G\,\hat a^\dag_I \hat a_I+(G-1)\,\hat a_R \hat a_R^\dag\approx0$. A threshold strategy can be easily defined, showing a $3$~dB advantage of the QBC over BC in the PAM and BPSK cases, where $|\langle\hat O_{PA}\rangle|$ is maximal. Indeed, if we consider the sample mean $\bar O_{PA}=\sum_{k=1}^M O_{PA}^k/{M}$, where $O_{PA}^k$ is the measurement outcome of the $k$-th copy, we declare the symbol $\{\bar \eta,\bar \phi\}=\argmin_{\{\sqrt{\eta},\phi\}\in \mathcal{A}}\left|\bar O_{PA}-\sqrt{\eta}e^{-i\phi}\right|$. It was shown in~\cite{Guha2009,Sanz17}, with a Cramer-Chernoff theorem based argument, that 
\begin{equation}
    P_{\mathsf{err}}^{PA}\leq e^{-d_{\mathsf{PAM( BPSK)}}^2 N_S M/2N_Z}.
    \label{eq:PerrlbPA}
\end{equation}
The EPE is twice the one found in \eqref{eq:Perrlbclassical}, which corresponds to a $3$~dB gain. However, the PA receiver does not provide any gain in the QPSK scheme since the symbols are not aligned.

{\it SFG-RX}: It maps the problem to the one of discriminating between coherent states \cite{Zhuang2017}. The SFG circuit is described by the interaction Hamiltonian
\begin{equation*}
    H_I=\hbar g \sum_{m=1}^M \left(\hat b^\dag \hat a_{R_m}\hat a_{I_m}+\hat b \hat a_{R_m}^\dag \hat a_{I_m}^\dag\right),
\end{equation*}
where $\{\hat a_{R_m},\hat a_{I_m}\}_{m=1}^M$ are the modes corresponding to the different copies of the RX and the idler, $\hat b$ is initially in the vacuum state, and $g$ is the coupling parameter. If we assume the low-brightness conditions $n_R(t)\equiv \langle \hat a_{R_m}^\dag \hat a_{R_m}\rangle_t\ll1$, $n_I(t)\equiv\langle \hat a_{I_m}^\dag \hat a_{I_m}\rangle_t\ll1$ and $|C(t)|^2\equiv|\langle\hat a_{S_m} \hat a_{R_m}\rangle_t|^2\ll1$, then one can solve the dynamics in the qubit approximation, finding that the mode $\hat b$ is in a coherent state mixed with a weak thermal noise~\cite{Zhuang2017}:
\begin{align*}
    C(t) &=C(0)\cos(\sqrt{M}gt), \\
    n_R(t) &=n_R(0),\quad n_I(t)=n_I(0),\\
    b(t) &=-i\sqrt{M} C(0)\sin(\sqrt{M}gt),\\
    n_b(t) &=[M|C(0)|^2+n_I(0)n_R(0)]\sin^2(\sqrt{M}gt),
\end{align*}
where $n_I(0)n_R(0)$ term in the last equation is the aforementioned thermal noise contribution. If we let evolve the circuit for a time $t_l=l\pi/2\sqrt{M}g$, with $l$ positive odd integer, then the correlations between the RX modes $C(t)$ disappears in favour of the coherent state amplitude $b(t)$. This results in the following input-output relations:
\begin{align}
    \hat a_{R_m}(t_l)&=\sqrt{1+|C(0)|^2}\,\hat a_{R_m}(0)-C(0)\,\hat a_{I_m}^\dag, \label{out1}\\
    \hat a_{I_m}(t_l)&=\sqrt{1+|C(0)|^2}\,\hat a_{I_m}(0)-C(0)\,\hat a_{R_m}^\dag. \label{out2}
\end{align}
Notice that $n_R(0)\simeq N_Z\gg1$, which lets the low-brightness condition fall. This issue is solved by sending the modes $\hat a_{R_m}$ to a low-transmissivity beam splitter, obtaining 
\begin{IEEEeqnarray*}{ll}
    \hat a_{R_m,1}^{(1)} &=\sqrt{\tau}\,\hat a_{R_m}+\sqrt{1-\tau}\,\hat v_m^{(1)}, \quad \text{and}\\
    \hat a_{R_m,2}^{(1)} &=\sqrt{1-\tau}\,\hat a_{R_m}-\sqrt{\tau}\,\hat v_m^{(1)},
\end{IEEEeqnarray*}
where $\hat v_m^{(1)}$ is a vacuum mode and $\tau N_Z\ll1$. We then send $\hat a_{R_m,1}^{(1)}$ and $\hat a_{I_m}^{(1)}=\hat a_{I_m}$ as inputs of the SFG circuit, which generates the mode $\hat b^{(1)}$ in a coherent state $|\sqrt{\tau M}C(0)\rangle$ and the outputs $\{\hat {a'}_{R_m}^{(1)}, \hat {a'}_{I_m}^{(1)}\}$ according to Eqs.~\eqref{out1}-\eqref{out2}. We mix the mode $\hat {a'}_{R_m}^{(1)}$ with $\hat a_{R_m,2}^{(1)}$, obtaining 
\begin{IEEEeqnarray*}{ll}
    \hat a_{E}^{(1)} &=\sqrt{\tau}\,\hat a_{R_m,2}^{(1)}+\sqrt{1-\tau}\,\hat {a'}_{R_m}^{(1)}, \\
    \hat a_{R}^{(1)} &=\sqrt{\tau}\,\hat {a'}_{R_m}^{(1)}-\sqrt{1-\tau}\,\hat a_{R_m,2}^{(1)}.
\end{IEEEeqnarray*}
We iterate this process $K$ times, by sending as input of the SFG the modes $\hat a_{R_m,1}^{(k)}=\sqrt{\tau}\,\hat a_{R_m}^{(k-1)}+\sqrt{1-\tau}\,\hat v_m^{(k)}$ and $\hat a_{I_m}^{(k)}=\hat {a'}_{I_m}^{(k-1)}$. This generates, as output, the $\hat b^{(k)}$ modes in a coherent state $|\sqrt{\tau M} C(0) [1-\tau(1+N_Z)]^k\rangle$ embedded in a thermal environment with $N_SN_Z\ll1$ average number of photons, and the $\hat a_E^{(k)}$ modes in an thermal state with mean number of photons $n_E^{(k)}=n_b^{(k)}=\tau M |C(0)|^2 [1-\tau(1+N_Z)]^{2k}$. The number of cycles $K$ is chosen such that $\sum_{k=1}^K n_{b}^{(k)}/\sum_{k=1}^{\infty} n_{b}^{(k)}=1-\varepsilon$, for some $\varepsilon\ll1$.\footnote{As a further technical remark, in the Chernoff bound regime we have $M|C(0)|^2\gg1$. Therefore, in order to apply the qubit approximation, we need to divide the $M$ signal-idler pair of modes in $M/\tilde K$ subsets, such that $\tilde K|C(0)|^2\ll1$, and apply the RX to the $\tilde K$ pair of modes at a time~\cite{Zhuang2017}.} As the mean number of photons $n_E^{(k)}$ and $n_b^{(k)}$ are both zero for $C(0)=0$, one can test the different hypothesis by simply applying a two-mode squeezing operation (TMSO) before the SFG circuit. The latter allows to displace to zero the phase-sensitive correlations of one of the hypothesis.\footnote{\cite{Zhuang2017} applies also two TMSO after the SFG circuit. This allows to have in every cycle $n_E^{(k)}=n_b^{(k)}$, and both $n_E^{(k)}$ and $n_b^{(k)}$ homogeneous functions of the phase-sensitive correlations of the hypothesis that we are testing.} We discuss the performance in the different QBC schemes in the $N_Z\gg1$, $N_S\ll1$, $\tau N_Z\ll1$, $\varepsilon\ll1$ limit. 

\begin{figure}[t]
	\centering
	\includegraphics[width=\columnwidth]{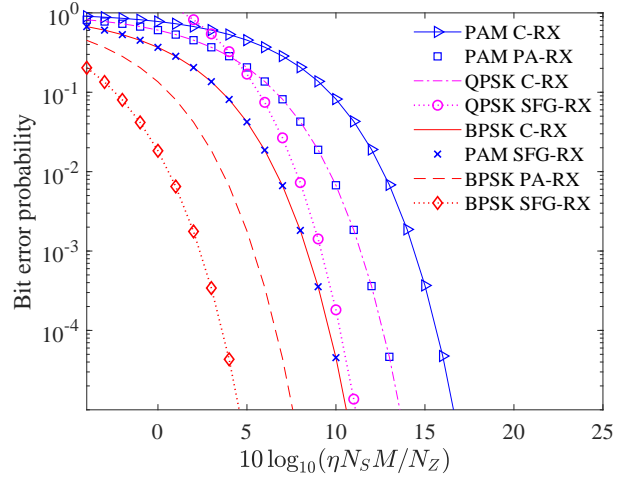} 
	\caption{Bit error probability for PAM, BPSK and QPSK for Heterodyne-RX in~\eqref{eq:Perrlbclassical}, PA-RX in~\eqref{eq:PerrlbPA} and SFG-RX in~\eqref{eq:Pe_PAM_Zhuang}-\eqref{eq:Pe_QPSK_Zhuang} as a function of $\eta N_s M/N_Z$.} 
	\label{fig:BER}
\end{figure}

\emph{a) PAM:} The SFG-RX achieves a $6$ dB gain in the EPE with respect the classical BC. We can simply apply a TMSO before the SFG, in order to have $n_E^{(k)}=n_b^{(k)}=0$ for the hypothesis $(\sqrt{\eta_1},0)$. One can simply measure the total number of photons $\sum_{k=1}^K (\hat n_b^{(k)}+\hat n_{E}^{(k)})$ and declare the aforementioned hypothesis if no photons are counted, obtaining the EP bound  in~ \eqref{eq:Pe_PAM_Zhuang}.

\emph{b) BPSK:} The same can be done here, by applying a TMSO before the SFG, in order to have $n_E^{(k)}=n_b^{(k)}=0$ for the hypothesis $(\sqrt{\eta},\pi)$. Then, we proceed like in the PAM case, achieving an EP given in \eqref{eq:Pe_BPSK_Zhuang}, which corresponds to a $6$ dB gain in the SNR with respect the classical case.

\emph{c) QPSK:} The task is equivalent to discriminate between the coherent states $|e^{i\phi} \sqrt{\sum_{k=1}^K (n_b^{(k)}+n_E^{(k)})}\rangle$, with $\phi\in\{0,\pi/2,\pi,3\pi/2\}$, where the TMSOs have the role of the displacements. This can be done following Bayesian strategies, like in the Dolinar receiver~\cite{Dolinar1973}. In alternative, we can simply test the different hypothesis by applying TMSOs, and discarding them as soon as one photon is detected~\cite{VanEnk02}, achieving the same $3$ dB gain in the SNR as in the Dolinar receiver, see in \eqref{eq:Pe_QPSK_Zhuang}. This is in contrast with the PA-RX, which does not provide any gain in this case. 
\begin{IEEEeqnarray}{ll}
    P_{\mathsf{err}}^\mathsf{PAM} & \leq e^{-d_\mathsf{PAM}^2N_S M/N_Z}, \IEEEyessubnumber \label{eq:Pe_PAM_Zhuang}\\
    P_{\mathsf{err}}^\mathsf{BPSK} & \leq e^{-4\sum_{k=1}^K (n_b^{(k)}+n_E^{(k)})}\simeq e^{-d_\mathsf{BPSK}^2 N_SM/N_Z},\IEEEyessubnumber \label{eq:Pe_BPSK_Zhuang}\\
    P_{\mathsf{err}}^\mathsf{QPSK} & \leq 4 e^{-\sum_{k=1}^K (n_b^{(k)}+n_E^{(k)})}\simeq 4e^{-d_\mathsf{QPSK}^2 N_SM/2N_Z}.\IEEEyessubnumber \label{eq:Pe_QPSK_Zhuang}
\end{IEEEeqnarray}

Figure \ref{fig:BER} illustrates the bit error probability (or interchangeably bit error rate) performance of Heterodyne, PA and SFG RXs for PAM, BPSK and QPSK modulations. In PAM, we have assumed that $\eta_2=\eta$ and $\eta_1=0$ corresponding to on-off-keying (OOK). The plots verify the results addressed in out analysis. For PAM and BPSK, the PA-RX and SGF-RX provide a 3 dB and a 6 dB EPE gain over the classical RX, respectively. For QPSK, only the SGF-RX provides a 3 dB EPE gain over the classical RX.

\section{QI-secured backscatter communication}\label{Sec:security}
This section discusses the possibility of using QI to enhance the secure backscatter communication.

The use of QI can provide a factor 4 gain in the error exponent for the communication link (referred to as Bob-Alice) between the backscatter device Bob and the transceiver Alice compared to the link (referred to as Bob-Eve) between Bob and the eavesdropping receiver Eve. This is due that the QI is not available on Bob-Eve link. In order for Eve to compensate this, Eve would need to use higher-gain antennas than that which Alice uses or be closer to Bob. Further, in order to increase the security that is to make the life of active eavesdropper more complicated, Alice can apply random phase shift to both the signal and idler paths. In case of BPSK, Eve would need to estimate this random phase shift before decoding the message from Bob such that the security of Bob-Alice link is enhanced. 

The system is vulnerable to active Eavesdropping attack where Eve illuminates the Bob's antenna with its own signal. Eve's signal will cause interference at Alice's receiver and could be detected. Adding power detector to Bob would also be utilized to detect Eve, but in practice this would mean that fraction of the power impinging at it's antenna would need to be fed to the detector thus reducing $\sigma_Q$ (and thus $\eta$). For instance, $50-50$ power divider at Bob's antenna would reduce the error exponent by factor 2 so that in case of PA-receiver would negate the gain achieved by using QI.

\section{Conclusions}
We proposed to use the quantum radar technology to enhance the performance of backscatter communication systems. Especially, we the use of quantum illumination to enhance the system performance.  The proposed QBC concept was verified to be within the reach of engineering applications such that in the PAM and BPSK cases, the quantum setting allows for a $6$ dB advantage in the error probability exponent while in the QPSK scheme a $3$ dB gain can be achieved using the SFG-RX. Finally, we would like to notice that in any scheme involving only phase modulation, as in the BPSK and QPSK protocols, the quantum setting allows for secure communication by means of quantum cryptography in a way similar to optical systems \cite{Shapiro2014}. 


\ifCLASSOPTIONcaptionsoff
  \newpage
\fi

\bibliographystyle{IEEEtran}

%

\end{document}